\input{aipcheck}

\documentclass[
    ,final            
  ]
  {aipproc}

\layoutstyle{6x9}

\usepackage{amsmath}

\newcommand{\bra}[1]{\langle#1|}
\newcommand{\ket}[1]{|#1\rangle}
\newcommand{\braket}[2]{\langle#1|#2\rangle}

\newcommand{\Tr}{\operatorname{Tr}}

\begin{document}

\title{How much a Quantum Measurement is Informative?}



\author{Michele Dall'Arno}{ address={Graduate School of Information
    Science, Nagoya University, Nagoya, 464-8601, Japan\\ICFO-Institut
    de Ciencies Fotoniques, E-08860 Castelldefels (Barcelona),
    Spain\\Quit group, Dipartimento di Fisica, via Bassi 6, I-27100
    Pavia, Italy} }

\author{Giacomo Mauro D'Ariano}{ address={Quit group, Dipartimento di
    Fisica, via Bassi 6, I-27100 Pavia, Italy\\Istituto Nazionale di
    Fisica Nucleare, Gruppo IV, via Bassi 6, I-27100 Pavia, Italy} }

\author{Massimiliano F. Sacchi}{ address={Quit group, Dipartimento di
    Fisica, via Bassi 6, I-27100 Pavia, Italy\\Istituto di Fotonica e
    Nanotecnologie (INF-CNR), P.zza L. da Vinci 32, I-20133, Milano,
    Italy } }

\begin{abstract}
  The informational power of a quantum measurement is the maximum
  amount of classical information that the measurement can extract
  from any ensemble of quantum states. We discuss its main
  properties. Informational power is an additive quantity, being
  equivalent to the classical capacity of a quantum-classical
  channel. The informational power of a quantum measurement is the
  maximum of the accessible information of a quantum ensemble that
  depends on the measurement. We present some examples where the
  symmetry of the measurement allows to analytically derive its
  informational power.
\end{abstract}

\maketitle


The information stored in a quantum system is accessible only through
a quantum measurement, and the postulates of quantum theory severely
limit what a measurement can achieve. The problem of evaluating how
much a measurement is informative has obvious practical relevance in
several contexts, such as the communication of classical information
over noisy quantum channels and the storage and retrieval of
information from quantum memories. When addressing such problem, one
can consider two different figures of merit: the probability of
correct detection~\cite{EE07} (in a discrimination scenario) and the
mutual information~\cite{DDS11} (in a communication scenario). The
latter case is the object of our discussion.

The {\em informational power} $W(\Pi)$ of a POVM $\Pi$ was introduced
in Ref.~\cite{DDS11} as the maximum over all possible ensembles of
states $R$ of the mutual information between $\Pi$ and $R$, namely
\begin{align*}
  W(\Pi) = \max_{R} I(R,\Pi).
\end{align*}

Informational power is an additive quantity. Given a channel $\Phi$
from an Hilbert space $\mathcal{H}$ to an Hilbert space $\mathcal{K}$,
the {\em single-use channel capacity} is given by $C_1(\Phi) := \sup_R
\sup_\Lambda I(\Phi(R),\Lambda)$, where the suprema are taken over all
ensembles $R$ in $\mathcal{H}$ and over all POVMs $\Lambda$ on
$\mathcal{K}$. A {\em quantum-classical channel}~\cite{Hol98} (q-c
channel) $\Phi_\Pi$ is defined as $\Phi_\Pi(\rho) := \sum_j \Tr[\rho
  \Pi_j] \ket{j}\bra{j}$, where $\Pi=\{\Pi_j\}$ is a POVM and
$\ket{j}$ is an orthonormal basis. In Ref.~\cite{DDS11} it was proved
that the informational power of a POVM $\Pi$ is equal to the
single-use capacity $C_1(\Phi_\Pi)$ of the q-c channel $\Phi_\Pi$,
i. e.
\begin{align*}
  W(\Pi) = C_1(\Phi_\Pi).
\end{align*}
The additivity of the informational power follows from the additivity
of the single-use capacity of q-c channels.

\begin{figure}[htb!]
  \centerline{\includegraphics[height=0.1\textheight]{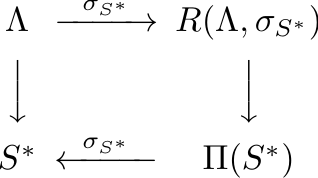}}
  \caption{The commuting diagram makes clear the duality between
    informational power and accessible information. Here $ S^* = \arg
    \max_S I(S, \Lambda)$ and $\Pi(S^*) = \arg \max_\Pi
    I(R(\Lambda,\sigma_{S^*}), \Pi)$. Horizontal arrows correspond to
    the duality operation of Eqs.~\eqref{eq:dualens}
    and~\eqref{eq:dualpovm}. Moving in the sense of the arrow
    corresponds to apply Eq.~\eqref{eq:dualpovm}, thus requiring
    $\sigma_{S^*}$. Moving in the opposite sense corresponds to apply
    Eq.~\eqref{eq:dualens}. The vertical arrow from $\Lambda$ to $S^*$
    indicates that $S^*$ is maximally informative for the POVM
    $\Lambda$, whereas the vertical arrow from
    $R(\Lambda,\sigma_{S^*})$ to $\Pi(S^*)$ indicates that $\Pi(S^*)$
    is maximally informative for the ensemble
    $R(\Lambda,\sigma_{S^*})$.  }
  \label{fig:commdiag}
\end{figure}

The informational power of a quantum measurement is equal to the
accessible information of a quantum ensemble that depends on the
measurement. The {\em accessible information}~\cite{Hol73} $A(R)$ of
an ensemble $R=\{p_i,\rho_i\}$ is the maximum over all possible POVMs
$\Pi$ of the mutual information between $R$ and $\Pi$, namely $A(R) =
\max_{\Pi} I(R,\Pi)$. Given an ensemble $S = \{ q_i, \sigma_i \}$,
define~\cite{Hal97Bus07BH09} the POVM $\Pi(S)$ as
\begin{align}
  \label{eq:dualens}
  \Pi(S):= \left\{ q_i \sigma_S^{-1/2} \sigma_i \sigma_S^{-1/2}
  \right\}.
\end{align}
Given a POVM $\Lambda = \{ \Lambda_j \}$ and a density matrix
$\sigma$, define~\cite{Hal97Bus07BH09} the ensemble $R(\Lambda,
\sigma)$ as
\begin{align}
  \label{eq:dualpovm}
  R(\Lambda, \sigma) := \left\{ \Tr[ \sigma \Lambda_j ], \frac {
    \sigma^{1/2} \Lambda_j \sigma^{1/2} } { \Tr[ \sigma \Lambda_j ]}
  \right\}.
\end{align}
In Ref.~\cite{DDS11} it was proved that the informational power of a
POVM $\Lambda = \{ \Lambda_j \}$ is given by
\begin{align*}
  W(\Lambda) = \max_\sigma A(R(\Lambda,\sigma)).
\end{align*}
Moreover, the ensemble $S^* = \{ q_i^*, \sigma_i^* \}$ is maximally
informative for the POVM $\Lambda$ if and only if
$\sigma_{S^*}=\arg\max_\sigma A(R(\Lambda,\sigma))$ and the POVM
$\Pi(S^*)$ is maximally informative for the ensemble
$R(\Lambda,\sigma_{S^*})$, as illustrated in the commuting diagram in
Fig.~\ref{fig:commdiag}.
From this result it immediately follows that for any given
$D$-dimensional POVM there exists a maximally informative ensemble of
$M$ pure states, with $D \le M \le D^2$, a result similar to Davies
theorem for accessible information~\cite{Dav78}. For POVMs with real
matrix elements~\cite{SBJOH99}, the above bound can be strengthened to
$D \le M \le D(D+1)/2$.

For any $D$-dimensional POVM $\Pi = \{ \Pi_j \}_{j=1}^N$ with
commuting elements, there exists a maximally informative ensemble $R =
\{p_i^*, \ket{i}\bra{i} \}_{i=1}^M$ of $M \le D$ states, where
$\ket{i}$ denotes the common orthonormal eigenvectors of $\Pi$, and
the prior probabilities $p_i^*$ maximize the mutual information. This
result applies to the problem of the purification of noisy quantum
measurements~\cite{DDS10}.

\begin{figure}[htb!]
  \centerline{\includegraphics[height=0.2\textheight]{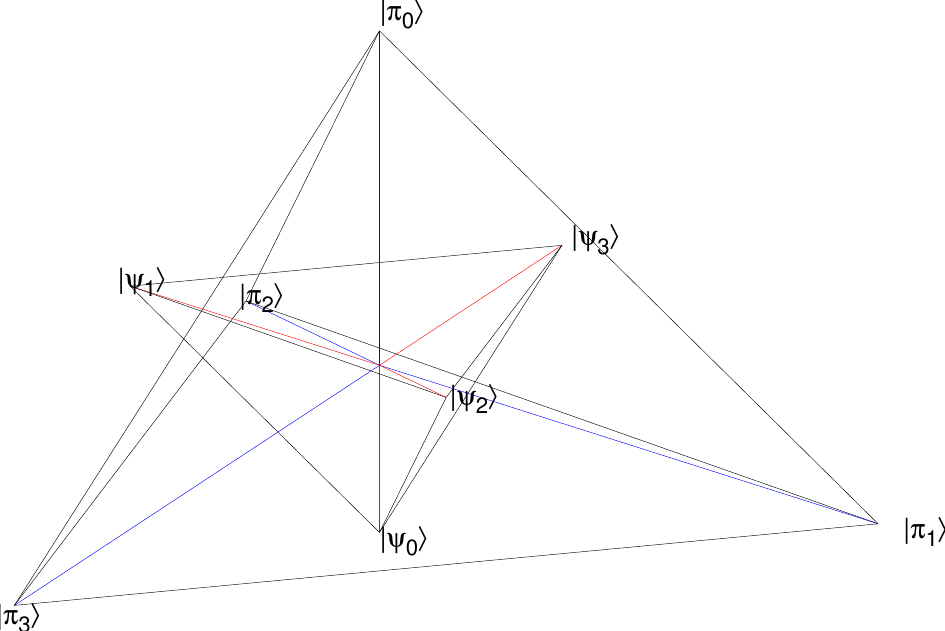}}
  \caption{(Color online) Representation of the $2$-dimensional SIC
    POVM (tetrahedral POVM, blue vectors) $\Pi =\{ \frac12
    \ket{\pi_j}\bra{\pi_j} \}_{j=0}^3$ an the corresponding maximally
    informative ensemble (anti-tetrahedral ensemble, red vectors) $R =
    \{ \frac14, \ket{\psi_i}\bra{\psi_i} \}_{i=0}^3$. Orthogonal
    states form an angle of $\pi$ as in the Bloch sphere
    representation, but their length is rescaled according to their
    norm.}
  \label{fig:sicpovm}
\end{figure}

For some class of $2$-dimensional and group-covariant POVMs (namely,
SIC POVMs~\cite{Dav78}, real-symmetric POVMs~\cite{SBJOH99},
mirror-symmetric POVMs~\cite{Fre06}), it is possible to provide an
explicit form for a maximally informative ensemble which enjoys the
same symmetry. For example (see Fig.~\ref{fig:sicpovm}), when $\Pi =\{
\frac12 \ket{\pi_j}\bra{\pi_j} \}_{j=0}^3$ is the $2$-dimensional SIC
POVM (tetrahedral POVM), the ensemble $R = \{ \frac14,
\ket{\psi_i}\bra{\psi_i} \}_{i=0}^3$ (anti-tetrahedral ensemble) with
$\braket{\pi_i}{\psi_i} = 0$ for any $i$ is maximally informative, and
the informational power is $W(\Pi) = \log\frac{4}{3}$.

The results we presented have obvious relevance in the theory of
quantum communication and measurement, and interesting related
works~\cite{OCMB11, Hol12} recently appeared. In particular, in
Ref.~\cite{Hol12} Holevo extends the results we presented to the
relevant infinite dimensional case.


\begin{theacknowledgments}
  This work was supported by JSPS (Japan Society for the Promotion of
  Science) Grant-in-Aid for JSPS Fellows No. 24-0219, Spanish project
  FIS2010-14830, Italian project PRIN 2008, and European projects
  COQUIT and Q-Essence.
\end{theacknowledgments}



\bibliographystyle{aipproc}   


\IfFileExists{\jobname.bbl}{}
 {\typeout{}
  \typeout{******************************************}
  \typeout{** Please run "bibtex \jobname" to optain}
  \typeout{** the bibliography and then re-run LaTeX}
  \typeout{** twice to fix the references!}
  \typeout{******************************************}
  \typeout{}
 }


\end{document}